\documentstyle [12pt] {article}
\newcommand{\h}[1]{\mathop{\lambda}\limits_{#1}\ \!\!\!}

\newcommand{\edf}{\ {\mathop{=}\limits^{\rm def}}\ }

\newcommand{\al}{\alpha}

\begin{document}
\begin{center}
\bf {ON THE RELATION BETWEEN MASS AND CHARGE:}\\
\bf {A Pure Geometric Approach.}
\end{center}
\begin{center}
\bf{M.I.Wanas\footnote{Astronomy Department, Faculty of Science,
Cairo University, Giza, Egypt.

E-mail:wanas@frcu.eun.eg}}
\end{center}

\begin{abstract}
 A new solution of the field equations of the generalized field
theory, constructed by Mikhail and Wanas in 1977, has been obtained.
 The geometric structure used, in the present
application, is an absolute parallelism (AP)-space with spherical
symmetry (type FIGI). The solution obtained represents a generalized
field outside a charged massive central body. Two schemes have been
used to get the physical meaning of the solution: The first is
related to the metric of the Riemannian space associated with the
AP-structure. The second is connected to a covariant scheme known as
{\it{Type Analysis}}. It is shown that the dependence on both
schemes for interpreting the results obtained, is more better than
the dependence on the metric of the Riemannian space associated with
the AP-structure.

In General, if we consider the solution obtained as representing a
geometric model for an elementary charged particle, then the results
of the present work can be summarized in the following points. (i)
It is shown that the mass of the particle is made of two
contributions: The first is the gravitational contribution, and the
second is the contribution due to the existence of charge. (ii) The
model allows for the existence of a charged particle whose mass is
completely electromagnetic in origin. (iii) The model prevents the
existence of a charged massless particle. (iv) The electromagnetic
contribution, to the mass, is independent of the sign of the
electric charge. (v) It is shown that the mass of the electron (or a
positron) is purely made of its charge.
\end{abstract}
\section{Introduction}
  In field theories, solutions with spherical symmetry have
received much attention, since they simplify several physical and
astrophysical problems. In the theory of general relativity, these
solutions may be used as models for static (or slowly rotating)
celestial objects with no net electric charge. While in
Einstein-Maxwell's theory they represent models for static objects
with non-vanishing net electric charge.

  The formation of a stable charged particle, in the framework of
classical electrodynamics, is impossible. This is due to the fact
that the field of an electric charge with a certain sign is
repulsive. Many authors believe that the best classical description
of elementary particles is that of singularities in a combined
electromagnetic-gravitational field (cf. [1]). Others have the hope
to explain the masses of charged particles by reducing them to the
energy of their electromagnetic fields [2]. This may lead to some
developments in the theory of elementary particles. A comprehensive
theory of gravity and electromagnetism may be very useful in
discussing this problem, since the existence of a gravitational
field (attractive) together with an electromagnetic field
(repulsive) may lead to a stable charged particle.

  The main problem with Einstein-Maxwell's theory is that the gravitational
field and the electromagnetic field are treated asymmetrically.
The gravitational field has been geometrized while the
electromagnetic field is introduced as a vector field added to the
symmetric tensor field $ g_{\mu\nu}$ of Riemannian geometry .
There is no definite role for the added field in the original
geometry.

 In the generalized field theory (GFT) [3], constructed by Mikhail and Wanas
in 1977,using Absolute Parallelism (AP)-geometry, the two fields
have been treated, in a comprehensive way, as one entity. All
geometric elements, used to describe gravitational and
electromagnetic quantities, are derived directly from the geometric
structure used. The theory has been applied to spherically symmetric
cases [4],[5], [6]. The solutions obtained, so far, were found to be
in agreement with previously known results, under some limiting
(particular) conditions.

  The aim of the present work is to explore a further solution, with
spherical symmetry. This may throw some light on the previously
stated problems. In section 2 the bases of the geometric structure
used for application is presented. A brief review of the GFT, the
version of AP-geometry used in its construction and a summary of its
applications are given in section 3. A new solution of the field
equation, of the GFT, is obtained in section 4. Some of the
electromagnetic  and  gravitational quantities are evaluated in
section 5. The results are discussed in section 6 and the paper is
concluded in section 7.
\section{ Summary of a Version of AP-geometry }
   In the present section, a brief review of the conventional AP-
   geometry, first used by Einstein [7] and developed by Robertson
   [8] and Mikhail [9], is given.
    The geometry used to construct the GFT is
 a version AP-geometry. An AP-space is an
affinely connected space, each point of which is labeled by a set of
n-independent variables $x^{\nu} (\nu =1,2,3,....n)$. At each point
we  introduce n-linearly independent contravariant vectors $
\h{i}^{\nu} (\ i =1,2,3,.....n)$. We denote by $\h{i}_{\mu}$ the
normalized co-factor of $\h{i}^{\mu}$ in the determinant of
$\h{i}^{\mu}$ so that,  $$
 \h{i}^\mu\h{i}_\nu=\delta^{\mu}_{\nu}~~~, \eqno{(1)}
 $$
 $$
\h{i}^\mu\h{j}_\mu=\delta_{\ i\ j}~~~, \eqno{(2)}$$ where summation
convention is carried out over repeated indices, regardless of their
positions. The affine connection $\Gamma^{\alpha}_{\mu\nu}$
characterizing the space will be defined as a consequence of
absolute parallelism condition,
$$ \h{i}^\mu_{+|~\nu}=0~~~, \eqno{(3)}$$ where the (+) sign and the stroke used
to characterize this type of tensor differentiation using the
 affine connection $\Gamma^{\alpha}_{\mu\nu}$ . For a
space of dimension 4 , there are 64 equations (3) with the unique
solution, $$
\Gamma^{\alpha}_{\mu\nu}=\h{i}^{\alpha}\h{i}_{\mu,~\nu}~~~,
\eqno{(4)} $$ where the comma is used for ordinary partial
differentiation. The curvature tensor corresponding to this
connection,
$$
{M}^{\al}_{.~\mu\nu\sigma}\edf {\Gamma}^{\al}_{.~\mu\sigma,\nu}-
{\Gamma}^{\al}_{.~\mu\nu,\sigma}+
{\Gamma}^{\epsilon}_{.~\mu\sigma}{\Gamma}^{\al}_{.~\epsilon\nu}-{\Gamma}^{\epsilon}_{.~\mu\nu}
{\Gamma}^{\al}_{.~\epsilon\sigma},\eqno{(5)}
$$
vanishes identically, while that corresponding to its dual
$\tilde\Gamma^{\alpha}_{\mu\nu}(= \Gamma^{\alpha}_{\nu\mu})$does not
vanish (a detailed geometric study of the curvature tensors in the
AP-space has been recently given  [10]). Tensor differentiation
using $\tilde\Gamma^{\alpha}_{\mu \nu}$ is distinguished by a
(-)sign. The torsion tensor, of the connection (4), is usually
defined by,
$$\Lambda^{\alpha}_{.~ \mu \nu} \edf\
{\Gamma^{\alpha}_{.~ \mu\nu}-\Gamma^{\alpha}_{.~ \nu\mu}}~~~.
\eqno{(6)}
$$ Another
third order tensor (contortion) is defined by, $$
\gamma^{\alpha}_{.~\mu \nu} \edf \h{i}^{\alpha} \h{i}_{\mu ; \nu}.
\eqno{(7)}, $$ where the semicolon is used to characterize covariant
differentiation using Christoffel symbol. The two tensors are
related by the formula,
$$\gamma^{\alpha}_{.\mu \nu}= \frac{1}{2} (\Lambda^{\alpha}_{.\mu
\nu } - \Lambda^{~ \alpha}_{\nu .\mu} - \Lambda^{~\alpha}_{\mu
.\nu}). \eqno{(8)}
 $$ A basic vector could be
 obtained by contraction of one of the above third order tensors,
 i.e.
$$ C_{\mu} \edf  \Lambda^{\alpha}_{.\mu \alpha }=
\gamma^{\alpha}_{. \mu \alpha}. \eqno{(9)}$$

Using the building blocks of the AP-space one can define the
following second order symmetric tensor,
$$g_{\mu\nu}\edf\h{i}_\mu\h{i}_{\nu}~~~.~~~\eqno{(10)}$$
This tensor is used as a metric tensor of the Riemannian space,
associated with the AP-space, when needed.

The following table contains second order tensors, defined in the
AP-space, that are used in most physical applications. It has been
originally constructed by Mikhail [7].
\newpage
\begin{center}
 Table 1: Second Order World Tensors [7]      \\
\vspace{0.5cm}
\begin{tabular}{|c|c|} \hline
 & \\
Skew-Symmetric Tensors                &  Symmetric Tensors   \\
 & \\ \hline
 & \\
${\xi}_{\mu \nu} \edf \gamma^{~ ~ \alpha}_{\mu \nu .
|{\stackrel{\alpha}{+}}} $ &
\\

${\zeta}_{\mu\nu} \edf C_{\alpha}~{\gamma^{~~ \alpha}_{\mu \nu .}
} $ &
\\
 & \\ \hline
 & \\
${\eta}_{\mu \nu} \edf C_{\alpha}~{\Lambda^{\alpha}_{.\mu \nu} } $
& ${\phi}_{\mu \nu} \edf C_{\alpha}~\Delta^{\alpha}_{.\mu \nu} $
\\

${\chi}_{\mu \nu} \edf \Lambda^{\alpha}_{. \mu
\nu|{\stackrel{\alpha}{+}} }$ & ${\psi}_{\mu \nu} \edf
\Delta^{\alpha}_{. \mu \nu|{\stackrel{\alpha}{+}}} $
\\

${\varepsilon}_{\mu \nu} \edf C_{\mu | {\stackrel{\nu}{+}}} -
C_{\nu | {\stackrel{\mu}{+}}}$ & ${\theta}_{\mu \nu} \edf C_{\mu |
{\stackrel{\nu}{+}}} + C_{\nu | {\stackrel{\mu}{+}}}  $
\\

${\kappa}_{\mu \nu} \edf \gamma^{\alpha}_{. \mu
\epsilon}\gamma^{\epsilon}_{. \alpha \nu} - \gamma^{\alpha}_{. \nu
\epsilon}\gamma^{\epsilon}_{. \alpha \mu}$   & ${\varpi}_{\mu \nu}
\edf  \gamma^{\alpha}_{. \mu \epsilon}\gamma^{\epsilon}_{. \alpha
\nu} + \gamma^{\alpha}_{. \nu \epsilon}\gamma^{\epsilon}_{. \alpha
\mu}$ \\
 & \\ \hline
 & \\
                  &  ${\omega}_{\mu \nu} \edf \gamma^{\epsilon}_{. \mu \alpha}\gamma^{\alpha}_{. \nu \epsilon}$   \\

                                      &  ${\sigma}_{\mu \nu} \edf \gamma^{\epsilon}_{. \alpha \mu} \gamma^{\alpha}_{. \epsilon \nu}$   \\

                                      &  ${\alpha}_{\mu \nu} \edf C_{\mu}C_{\nu}$   \\

                                      &  $R_{\mu \nu} \edf \frac{1}{2}(\psi_{\mu \nu} - \phi_{\mu \nu} - \theta_{\mu \nu}) + \omega_{\mu \nu}$          \\
 & \\ \hline
\end{tabular}
\end{center}
$$
where \ \ \ \Delta^\alpha_{. \mu \nu} \edf \gamma^\alpha_{. \mu \nu}
+ \gamma^\alpha_{. \nu \mu}.
$$

\section{A Brief Review of GFT: Bases and Results}
The Generalized Field Theory is a pure geometric attempt to unify
gravity and electromagnetism. It has been constructed in 1977 by
Mikhail and Wanas [3]. In general, the theory is a covariant one.
Its anti-symmetric section, representing electromagnetism in the
weak field regime, is gauge invariant. The underlaying geometry, for
this theory, is a version of the AP geometry. This version is
slightly different from that usually used in the literature, since
it is based on the dual connection and is of the Riemann-Cartan
type. The main advantages of this  version are clarified in the
following lines. Consider the tensor derivative, of an arbitrary
covariant vector $A_\mu$, using the dual connection,
$$
A_{\stackrel\mu{-}|\nu}\edf A_{\mu,\nu}- A_{\al}
\tilde\Gamma^{\al}_{.~\mu\nu}. \eqno{(11)}
$$
The commutation formula of this type of differentiation, for an
arbitrary covariant vector, can be expressed by the following
relation,
$$
A_{\stackrel\mu{-}|\nu\sigma}-A_{\stackrel\mu{-}|\sigma\nu}=A_{\al}
\tilde M^{\al}_{.~\mu\nu\sigma}-A_{\stackrel\mu{-}|\al}
~~\tilde\Lambda^{\al}_{.~\nu\sigma},\eqno{(12)},
$$
where,
$$
\tilde{M}^{\al}_{.~\mu\nu\sigma}\edf
\tilde{\Gamma}^{\al}_{.~\mu\sigma,\nu}-
\tilde{\Gamma}^{\al}_{.~\mu\nu,\sigma}+
\tilde{\Gamma}^{\epsilon}_{.~\mu\sigma}\tilde{\Gamma}^{\al}_{.~\epsilon\nu}-\tilde{\Gamma}^{\epsilon}_{.~\mu\nu}
\tilde{\Gamma}^{\al}_{.~\epsilon\sigma},\eqno{(13)}
$$
is the conventional curvature tensor and
$\tilde\Lambda^{\al}_{.~\nu\sigma}$ is the torsion tensor
corresponding to the dual connection. It is worth of mention that
both tensors are , simultaneously, non vanishing in this version.
This is another  difference between the present version and the
AP-version used in the literature. Because of this difference we
attribute, to this version of geometry, a Riemann-Cartan character.
Now if we replace the covariant vector in equation (12) by the
covariant tetrad vectors, then this equation can be written as:

$$\h{i}_{\stackrel{\mu}{-}|\nu\sigma}-\h{i}_{\stackrel{\mu}{-}|\sigma\nu}=\h{i}_{\al}
\tilde{M}^{\al}_{.~\mu\nu\sigma} - \h{i}_{\stackrel{\mu}{-}|\al}
\tilde\Lambda^\al_{. \nu \sigma},\eqno{(14)}$$ which can be written
in the alternative form:
$$\h{i}_{\stackrel{\mu}{-}|\nu\sigma}-\h{i}_{\stackrel{\mu}{-}|\sigma\nu}=\h{i}_{\al}
\tilde{L}^{\al}_{.~\mu\nu\sigma},\eqno{(15)}$$ where
$\tilde{L}^{\al}_{.~\mu\nu\sigma}$ is a fourth order tensor defined
by:
$$
\tilde{L}^{\al}_{.~\mu\nu\sigma}\edf\h{i}^{\al}(\h{i}_{\stackrel{\mu}{-}|\nu\sigma}
-\h{i}_{\stackrel{\mu}{-}|\sigma\nu}). \eqno{(16)}$$ We call this
tensor  {\it  the non-conventional curvature tensor}. It is a fourth
order tensor defined in the present version of the AP-geometry. It
is a unique geometric object which measures the non-commutativity of
covariant differentiation, using the dual connection. This tensor,
implicitly, contains both conventional curvature and torsion
tensors. It represents the second main difference between the
present version of AP-geometry and the conventional AP-version .

The non-conventional curvature (16) has been used to construct the
GFT [3]. The scalar Lagrangian function of the theory is obtained by
double contractions of this tensor. Using a certain variational
technique, the authors of the theory obtained the differential
identity,
 $$E^\mu_{\ . ~ \nu |{\stackrel{\mu}{-}}}=0. \eqno{(17)} $$
Considering this identity as representing a general conservation
law, the  field equations of the GFT have been written as,
$$E_{\mu\nu}=0~~~, \eqno{(18)}
$$ where,
$$E_{\mu\nu}\edf\ g_{\mu\nu}\L-2L_{\mu\nu}-2g_{\mu\nu}
C^\gamma_{-|~\gamma}-2C_{\mu} C_{\nu}- 2 g_{\mu\alpha} C^{\epsilon}
\Lambda^{\alpha}_{\ . ~\epsilon\nu}+2 C_{{\stackrel{\nu}{+}}|~\mu}-
2 g^{\gamma\alpha} \Lambda_{\mu\nu\alpha|{\stackrel{\gamma}
{+}}}~~~,~~~\eqno{(19)} $$ and
$$\L_{\mu\nu}\edf\Lambda^{\delta}_{\ . ~\epsilon\mu}
\Lambda^{\epsilon}_{\ . ~\delta\nu}-C_{\mu}C_{\nu}~~~.~~~
\eqno{(20)}$$  This shows that $E_{\mu\nu}$ is a non-symmetric
second order tensor defined in the AP-space.

In the following lines, the results of applications of the GFT are
summarized. \\
(i) In the limits of  weak static fields and slowly moving test
particles [11], the theory gives rise to both Newton's gravity and
Maxwell's electromagnetism. Also, calculations in this frame shows
that Lorentz condition, a condition usually used to fix the gauge in
the solutions of Maxwell's equations, is not imposed from outside
the GFT , but it arises very naturally as a result of the presence
of gravity. The natural appearance of this condition represents a
type
of interaction between gravity and electromagnetism.\\
(ii) Using an AP-structure with spherical symmetry, of the type F0GI
representing the absence of electromagnetism, the metric of the
Riemannian space associated with the solution, of the GFT field
equations, is found to be
identical to the Schwarzschild exterior metric [4].\\
(iii) In an application [5], using an AP-structure of the type FIGI,
the metric of the Riemannian space associated with the solution
obtained is found to be similar to Reissner-Nordstrom solution with
an extra term. This term tends to zero  at large enough distances
from the source of the field. In that application, if the constant
of integration assigned to the electric charge of the source
vanishes, then the metric will reduce to the Schwarzschild metric
and the type of the space will be F0GI.\\
(iv) In a cosmological application of the GFT [12], using an
AP-structure of the type F0GIII (representing the absence of
electromagnetism and the presence of a strong gravitational field),
a unique solution of the field equations has been obtained. This
solution represents a pure geometric world model. It is a non-empty
expanding model with a density parameter $= 0.75$. The model fixes
the value of the curvature constant $k = -1$, and is free from
particle horizons.\\
(v) Another application of a solution of the type FIGI, having
spherical symmetry, is carried out [6]. This solution shows a new
interaction between gravity and electromagnetism. It shows that, a
rotating neutral massive object can produce a global magnetic field.
The comparison between the theoretical results obtained and
astrophysical observations of the magnetic field of diverse
celestial objects shows good qualitative agreement.
\section{A New Solution Of The Field Equations }
The geometric structure to be used in the present application is an
AP-space having spherical symmetry with  n=4. The structure of this
space has been derived by Robertson [8].  The tetrad vector field
characterizing this geometric structure is given, in spherical polar
coordinates, by:
$$(\h{i}^{\mu})={\left(\matrix{
               A    &Dr                     &0
               &0\cr
               0    &B\sin\theta\cos\phi    &\frac{B}{r}\cos\theta\cos\phi
               &-\frac{B\sin\phi}{r\sin\theta}\cr
               0    &B\sin\theta\sin\phi    &\frac{B}{r}\cos\theta\sin\phi
               &\frac{B\cos\phi}{r\sin\theta}\cr
               0    &B\cos\theta            &-\frac{B}{r}\sin\theta           &0
}\right)}~~, \eqno{(21)}$$ where $A,B$ and $D$ are functions of $r$
only. It has been shown [5] that the geometric structure, given by
the matrix (21), has the type FIGI (see section 4). This means that
this structure is capable of representing a combined gravitational
and electromagnetic weak field outside a spherically symmetric
material distribution.

From (10) we get for (21), $$\left.
\begin{array}{lllllll}
g_{00}&=&\frac {B^{2}+D^{2} r^{2}}
{A^{2} B^{2}} &, &g_{01}&=&\frac {-D r} {A B^{2}}\vspace{2mm}\\
g_{11}&=&\frac {1} {B^{2}}~~,&g_{22}=\frac{r^{2}}{B^{2}} ~~~,
&g_{33}&=&\frac{r^{2}\sin^{2}\theta} {B^{2}}~~~.
\end {array}
 \hspace{1cm} \right\}\eqno{(22)} $$
   Evaluating $E_{\mu\nu}$ in the space (21),
and substituting into the field equations (18), we get the following
set of differential equations:
$$\frac{B^{2}+D^{2}r^{2}} {A^{2}}[b(r)+\frac{D^{2}r^{2}}
{B^{2}}\ell(r)]=0~~~,\eqno{(23)}$$
$$\frac{Dr}{A}[B(r)+\frac{D^{2}r^{2}}{B^{2}}\ell(r)]=0~~~,\eqno{(24)}$$
$$-\frac{B'}{B}^{2}+\frac{2}{r}(\frac{B'}{B}+\frac{A'}{A})-2\frac{A'B'}
{A B}-\frac{D^{2}r^{2}}{B^{2}} \ell(r)=0~~~,\eqno{(25)}$$
$$\frac{D^{2}r^{2}}{B^{2}}[-\ell(r)+\frac{A''}{A}-\frac{D''}{D}
+\frac{4}{r}(\frac{A'}{A}-\frac{D'}{D})-2(\frac{A'}{A})^{2}-(\frac{D'}{D})^{2}+3(\frac
{A'D'}{AD}-\frac{A'B'}{A B} +\frac{B'D'}{B D})] $$
$$+\frac{A''}{A}+\frac{B''}{B}-2(\frac{A'}{A})^{2}
-(\frac{B'}{B})^{2}+\frac{1}{r}(\frac{A'}{A}+\frac{B'}{B})=0
~~~,\eqno{(26)}$$ where, $$b(r)\edf\
3(\frac{B'}{B})^{2}-\frac{4}{r}\frac{B'}{B}-2\frac{B''}{B}
~~~,\eqno{(27)}$$ $$\ell(r)\edf \
5(\frac{B'}{B})^{2}-\frac{8}{r}\frac{B'}{B}-2\frac{B''}{B}+
\frac{2}{r}\frac{D'}{D}-2\frac{B'}{B}\frac{D'}{D}+\frac{3}{r^{2}}
~~~,\eqno{(28)}$$ and the $(')$ represents differentiation with
respect to $r$. For these differential equations we have previously
obtained three different solutions [5]. A fourth solution, of
physical interest, could be obtained as follows. The set of
equations (23) -(26) could be written in the more reduced form:
$$\frac{B^{2}+D^{2}r^{2}}{A^{2}}[b(r)+\frac{D^{2} r^{2}} {B^{2}}
\ell(r)]=0 \eqno{(29)}$$
$$\frac{Dr}{A}[b(r)+\frac{D^{2}r^{2}}{B^{2}}\ell(r)]=0\eqno{(30)}$$
$$f(r)-\frac{D^{2}r^{2}}{B^{2}}\ell(r)=0\eqno{(31)}$$
$$s(r)+\frac{D^{2}r^{2}}{B^{2}}K(r)=0\eqno{(32)}$$ where,
$$f(r)\edf\
-(\frac{B'}{B})^{2}+\frac{2}{r}(\frac{B'}{B}+\frac{A'}{A})
-2\frac{A'}{A} \frac{B'}{B}~~~,\eqno{(33)}$$
$$s(r)\edf\frac{A''}{A}+\frac{B''}{B}-2(\frac{B'}{B})^{2}+\frac{1}{r}
(\frac{A'}{A}+\frac{B'}{B})~~~,\eqno{(34)}$$ $$K(r)\edf\
-\ell(r)+\frac{A''}{A}-\frac{D''}{D}+\frac{4}{r}(\frac{A'}{A}
+\frac{D'}{D})-2(\frac{A'}{A})^{2}-(\frac{D'}{D})^{2}
+3(\frac{A'}{A}\frac{D'}{D}-\frac{A'}{A}\frac{B'}{B}+\frac{B'}{B}
\frac{D'}{D})\eqno{(35)}$$ We exclude the solution,
$$B^{2}+D^{2}r^{2}=0 ~~~,~~~ $$ since it will cause the appearance
of imaginary functions. We are interested in solutions that
represent combined electromagnetic-gravitational fields. The
vanishing of $D$ will change the type of space from FIGI to F0GI
which represents pure gravity. So, the solution, $$D=0$$, is to be
excluded as well. Now taking $b(r)=0$, the set of differential
equations (29)-(32) will  reduce to :$$ \left.
\begin{array}{llll}
       &b(r)                           &=    &0~~,\vspace{2mm}\\
       &\ell(r)                        &=    &0~~,\vspace{2mm}\\
       &f(r)                           &=     &0~~,\vspace{2mm}\\
 s(r)  &+\frac{D^{2}r^{2}}{B^{2}}K(r)  &=     &0~~,
 \end{array}
 \hspace{1cm} \right\} \eqno{(36)} $$

Using (27), the first equation of the set (36) could be integrated
twice to give:
$$B=\frac{1}{(\beta_{1}+\frac{\beta_{2}}{r})^{2}}~~~,
\eqno{(37)}$$ where $\beta_{1}$, $\beta_{2}$ are the constants of
integration. Substituting (37) in (28), the second equation of the
set (36) can be integrated to give
:$$D=\frac{\alpha}{r^{\frac{3}{2}}(\beta_{1}+\frac{\beta_{2}}{r})^{2}
(\beta_{1}-\frac{\beta_{2}}{r})}~~~.~~~\eqno{(38)}$$ where $\alpha$
is a constant of integration. Again using (37) and (33), the third
equation of the set (36) can be integrated to
give:$$A=\beta_{3}\frac{(\beta_{1}+\frac{\beta_{2}}{r})}{(\beta_{1}-\frac
{\beta_{2}}{r})}~~~.~~~ \eqno{(39)}$$ where $\beta_{3}$ is another
constant of integration.\\

  {\bf Boundary Conditions}: The AP-space (21) has spherical symmetry,
then it is natural to assume that
$\h{i}^\mu\rightarrow\delta_{i}^{\mu}$ (in Cartesian coordinates) as
$r\rightarrow\infty$. This could be achieved if
$$ \left.
\begin{array}{lll}
 A   &\rightarrow  &1 \vspace{2mm}\\
 B  &\rightarrow  &1\vspace{2mm}\\
 D  &\rightarrow  &0\,
 \end{array}
 \hspace{1cm} \right\} as~~~ r\rightarrow \infty. \eqno{(40)}
$$
This implies taking, $$\beta_{1}=\beta_{3}=1~~.\eqno{(41)}$$ For
later convenience we take~,$$\beta_{2}=\frac{1} {2}
m~~,\eqno{(42)}$$ where $m$ is constant. So, the full solution of
the set (36)can be written in the form:\\ $$ \left.
 \begin{array}{lll}
 A         &=       &\frac{(1+\frac{m}{2r})}{(1-\frac{m}{2r})}~~,\vspace{2mm}\\
 B         &=       &\frac{1} {(1+\frac{m} {2r})^{2}}~~, \vspace{2mm}\\
 D         &=
 &\frac{\alpha}{{r^{\frac{3}{2}}}(1+\frac{m}{2r})^{2}(1-\frac{m}{2r})}~~~.
 \end{array}
 \hspace{1cm} \right\}\eqno{(43)}$$ It can be shown that solution (43) satisfies
the fourth equation of the set (36) without any further conditions.
\section{Physical Interpretation}
     In geometric field theories, one is dealing all the time with
geometric objects. In order to attribute some physical meanings to
these objects, authors usually use the following scheme. They
compare the new theory with previous field theories, dealing with
the same interaction(s), using certain limiting conditions. For
example, the first  results supporting the general theory of
relativity (GR) have been obtained by comparing the theory with
Newton's theory of gravitation, using the conditions of weak static
field and slowly moving test particles. As mentioned above, the GFT
is an attempt to unify gravity and electromagnetism. It has been
compared [11] with both GR and Maxwell's theory of electromagnetism.
The results of this comparison have attributed some physical
meanings, summarized in Table 2, to certain geometric objects using
certain system of units.

\begin{center}
 Table 2: Physical Meaning of Geometric Objects.      \\
\vspace{0.5cm}
\begin{tabular}{|c|c|} \hline
 & \\
Geometric Object                &  Physical Meaning   \\
 & \\ \hline
 & \\
$g_{\mu \nu} $ & Gravitational Potential
\\

 & \\ \hline
 & \\
$C_\mu$ & Generalized electromagnetic Potential
\\ & \\
\hline & \\ $F_{\mu \nu}$   & Electromagnetic Field  Tensor
\\
 & \\ \hline
 & \\ $T_{\mu \nu}$
                  &  Material-Energy-Stress Tensor \\ & \\  \hline
\end{tabular}
\end{center}
where,
$$F_{\mu\nu} \edf Z_{\mu\nu} - \xi_{\mu\nu}, $$ $$Z_{\mu\nu} \edf
\eta_{\mu \nu} + \zeta_{\mu\nu },$$ $$ T_{\mu\nu} \edf
g_{\mu\nu}\Lambda + \varpi_{\mu\nu} - \sigma_{\mu\nu}, $$
$$\Lambda \edf \frac{1}{2}(\sigma - \varpi). $$

In addition to the scheme mentioned above, a new scheme called {\it
Type Analysis} has been suggested [11], in the context of the GFT,
to know the capabilities of an AP-structure to represent physical
fields. This procedure is usually applied before solving the field
equations. It is a covariant procedure since it depends on the
values of certain tensors. It can be used to, physically, classify
AP-structures to be used for applications (cf. [13] ). Table 3 lists
the values of the tensors admitted by a certain geometric
AP-structure, the physical meaning of such values and the
corresponding codes attributed to these values. It is to be stressed
that the first column of this table does not represent conditions
on, but values of, the corresponding tensors as calculated from a
given AP-structure. The suggested scheme has shown its advantages in
several applications (cf. [5], [6], [12]\& [14]).

\begin{center}
 Table 3: Type Analysis       \\
\vspace{0.5cm}
\begin{tabular}{|c|c|c|} \hline
& & \\ Tensor Values  & Physical Meaning  & Code
\\ & &
\\ \hline & & \\ $F_{\mu\nu}= 0$     & No electromagnetic field. & $F0$
\\ & & \\ \hline & & \\ $F_{\mu\nu} \neq 0 $, $Z_{\mu\nu} =0 $ &  Weak electromagnetic field.   &
$FI$ \\ & & \\ \hline & & \\ $F_{\mu\nu} \neq 0 $, $Z_{\mu\nu} \neq
0$  & Strong electromagnetic field. & $FII $\\ & & \\ \hline \hline
& &
\\ $R^{\alpha}_{.\beta\gamma\delta} = 0$  & No gravitational
field. & $G0$ \\ & & \\ \hline & & \\
$R^{\alpha}_{.\beta\gamma\delta} \neq 0$, $T_{\mu\nu} = 0$,
$\Lambda = 0$ & Weak gravitational field in free space. & $GI$ \\
& & \\ \hline & &\\ $R^{\alpha}_{.\beta\gamma\delta} \neq 0$,
$T_{\mu\nu} \neq 0$, $\Lambda = 0$ & Gravitational field within a
material distribution. & $GII$ \\ & & \\ \hline & & \\
$R^{\alpha}_{.\beta\gamma\delta} \neq 0$, $T_{\mu\nu} \neq 0$,
$\Lambda \neq 0$ & Strong gravitational field within a material
distribution. & $GIII$
\\ & & \\ \hline
\end{tabular}
\end{center}
The type of the space, under consideration, is written in two parts:
The first is one of the codes written in the first three rows of
table 3, carrying information about the capability of an
AP-structure to represent  electromagnetic fields. The second is one
of the codes written in the other rows of this table and carrying
information about the capability of the same structure to represent
gravitational fields.
\newpage

\subsection{Electromagnetic Quantities}
    In the generalized field theory , the vector $C_{\mu}$ has been
identified as the generalized electromagnetic potential [3]. The
non- vanishing components of this vector, in the space given by
(21), have been found to have the values [5]: $$C_{0}=\frac{Dr}
{A}(\frac{B'} {B}-\frac{D'} {D}-\frac{3} {r})~~~,\eqno{(44)}$$
$$C_{1}=\frac{A'} {A}+2\frac{B'} {B}~~~.~ \eqno{(45)}$$ The
electromagnetic field tensor $F_{\mu\nu}$ is connected to the
electromagnetic potential via the skew-symmetric part of the field
equation of the GFT (18),
$$F_{\mu\nu}=C_{\mu,\nu} -C_{\nu,\mu}~~~.~\eqno{(46)}$$ From
(45) and (46) it is clear that $C_{1}$ will not produce any
electromagnetic field, since $A$ and $B$ are functions of $(r)$
only. Then the only non- vanishing components of $F_{\mu\nu}$, in
the space (21), are
$$F_{01}=-F_{10}=C_{0,1}~~~~.~\eqno{(47)}$$ Using solution (43) to
evaluate (44), we get:$$C_{0}=\frac{\alpha} {2 r^{\frac{3} {2}}}
(\frac{1}{(1+\frac{m}{2r})^{2}(1-\frac{m}{2r})}-\frac{4}{(1+\frac{m}
{2r})^{3}})~~~,~~~\eqno{(48)}$$ and for (47) we
get:$$F_{01}=-F_{10}=\frac{\alpha} {4r^\frac{5} {2}(1+\frac{m}
{2r})^{4}(1-\frac{m} {2r})^{2}} [9-37(\frac{m} {2r})+35(\frac{m}
{2r})^{2}-15(\frac{m} {2r})^{3}]~~~.~~~\eqno{(49)}$$ It is clear
from(48) and (49) that the constant $\alpha$ has a direct relation
to the charge of the system. The vanishing of  $\alpha$  will lead
to the vanishing of the electromagnetic field. Also, the vanishing
of $\alpha$  will change the type of the space from FIGI to F0GI,
and this represents pure gravity , as stated above. It is to be
considered that (49) represents, in general,  the non-vanishing
components of a tensor, in a certain coordinate system.
Consequently, this tensor will never vanish in any other coordinate
system,  in the context of the new solution (43). This will be
discussed later.
\subsection{The Gravitational Potential}
    The symmetric tensor $g_{\mu\nu}$, defined by (10), has been
identified in a previous work [3] as the gravitational potential.
Using (10) and the solution (43) we get the following non-vanishing
components of the gravitational potential: $$ \left.
\begin{array} {llll}
 g_{00} & =  &\frac{(1-\frac{m} {2r})^{2}} {(1+\frac{m} {2r})^{2}}& +~~~~~\frac{\alpha^{2}} {r(1+\frac{m}{2r})^{2}}~~,\vspace{2mm}\\
 g_{01} & =  &g_{10}&= -\frac{\alpha}{r^\frac{1}{2}}(1+\frac{m}{2r}),\vspace{2mm}\\
 g_{11} & =  &(1+\frac{m} {2r})^{4} & ~~,\vspace{2mm}\\
 g_{33} & =  &g_{22}\sin\theta^{2}  &= g_{11} r^{2}\sin\theta^{2}.
\end{array}
\hspace{1cm} \right\}\eqno{(50)}$$
 To facilitate comparison with general relativity, concerning gravity,
we write the indefinite metric of Riemannian space associated with
the AP-space(21). This metric is defined in terms of the tetrad
vectors via the relation: $$\left.
\begin{array}{lll}
 ds^{2}       &=         &*g_{\mu\nu} dx^{\mu}dx^{\nu}\vspace{2mm}\\
 where~~~
 *g_{\mu\nu}  &\edf     &\sum_{i=1}^{4}e_{i}\h{i}_{\mu}\h{i}_{\nu}\vspace{2mm}\\
 and~~~
  e_{i}       &=        &(1, -1,-1,-1)~~.
  \end{array}
  \hspace{1cm}\right\} \eqno{(51)}$$
Using the solution (43), the metric (51) can be written in the form:
$$ds^{2}=[\frac{(1-\frac{m} {2r})^{2}} {(1+\frac{m}
{2r})^{2}}-\frac{\alpha^{2}} {r(1+\frac{m}
{2r})^{2}}]dt^{2}+\frac{2\alpha} {r^\frac{1}{2}}(1+\frac{m}{2r})dr
dt -(1+\frac{m} {2r})^{4} (dr^{2}+r^{2} d\theta^{2}+r^{2}
\sin\theta^{2} d\phi^{2})~~.~~\eqno{(52)}$$ It is of interest to
note that the vanishing of $\alpha$  will reduce the metric (52) to
the Schwarzschild metric in its isotropic form. This supports the
result obtained in the previous subsection (5.1) that $\alpha$ has
to be related to the electric charge of the system under
consideration.
\section{General Discussion}

    (i) It is shown in subsection 5.1 that the vanishing of the constant
$\alpha$  will give rise to the vanishing of the electromagnetic
field. Also it is shown in subsection 5.2 that the vanishing of
$\alpha$ reduces the metric of the associated Riemannian space to
the Schwarzschild form. Furthermore, as mentioned above, the
vanishing of $\alpha$ will change the type of the space (21) from
FIGI to F0GI. These three evidences are consistent with our
identification of $\alpha$, that it has to be related to the
electric charge of the system. From dimensional consideration,
$\alpha$ cannot be directly identified as the charge of the system.
Previous experience in Einstein-Maxwell's theory and in classical
electrodynamics indicates that: (a) The geometric charge (the charge
measured in relativistic units G=c=1) is measured in cm, (b)The
electromagnetic field strength and the electromagnetic potential are
both linear in the electric charge of the system. In the present
treatment it is clear from  (50) or (52) that $\alpha$ has the
dimension $cm^{\frac{1} {2}}$~~; while the electromagnetic field
(49) and potential (48) are both linear in $\alpha$ which is
consistent with (b). So, to get an agreement between the present
results and (a) we should take,
$$\alpha^{2}=\frac{\epsilon^{2}} {a}~~,~~ \eqno{(53)}$$ where
$\epsilon$ is the geometric charge of the system  and $a$  is some
constant length to be fixed later.

  (ii) Using the coordinate transformation:$$\left.
  \begin{array}{lll}
  T      &=       &t+\psi(r)~~~,\vspace{2mm}\\
 \psi(r) &=       &\int\frac{\alpha r^{\frac{1} {2}}(1+\frac{m}{2r})^{3}} {r (1-\frac{m} {2r})^{2}-\alpha^{2}}dr~~~,\vspace{2mm}\\
 R       &=       &r(1+\frac{m} {2r})^{2}~~~,
 \end{array}
 \hspace{1cm}\right\}\eqno{(54)}$$ we can
write the metric (52) in the following form: $$ds^{2}=\gamma
dT^{2}-\frac{dR^{2}} {\gamma}-R^{2} d\theta^{2}-R^{2} \sin\theta^{2}
d\phi^{2}\eqno{(55)}$$ where $$\left.
\begin{array}{lll}
 \gamma     &=     &1-\frac{2\mu} {R}~~~,\vspace{2mm}\\
 \mu        &=     &m+\frac{1} {2}\alpha^{2}~~~,\vspace{2mm}\\
            &=     &m+\frac{1} {2} \frac{\epsilon^{2}} {a}.
 \end{array}
 \hspace{1cm}\right\}\eqno{(56)}$$ The metric
(55), apparently, has the form of Schwarzschild exterior metric in
its standard form. This leads to the following interesting physical
meaning. The constant $\mu$ can be identified as the total mass of
the body causing the curvature of the space. This mass constitutes
two parts: The first is the mass $m$ which can be identified as the
gravitational mass of the body, while the second is a contribution
of the electric charge of the body to its mass. The second part is
independent of the sign of the electric charge , as clear from (56).

   (iii)Although the metric (55), apparently, has the Schwarzschild standard form ,
the model represents a combined electromagnetic-gravitational field.
This is supported by: (a) The non-vanishing of the electromagnetic
field tensor  $F_{\mu\nu}$  in the coordinate system
(t,r,$\theta$,$\phi$) as given by (49). The electromagnetic field
will never vanish in any other coordinate system, as far as one is
using solution (43), since it is represented by a tensor; (b) The
appearance of $\alpha^{2}$ term in the metric tensor (52), as
$\alpha$ being connected with the electromagnetic field as shown
above.

  The type of the space used has been found to be FIGI. It is to be
considered that the {\it type analysis}, by which we know the type
of the space, is a procedure independent of the coordinate system
used , as it depends completely on tensors of different types. In
other words, the type of the space is invariant under general
coordinate transformation.

   (iv)Another interesting result could be obtained from (56). It is
shown that as \\ $\epsilon\neq{0}$ ~,~ $\mu$  will not vanish. On
the other hand, if  $\mu=0$ , $m$ will be negative which cannot be
accepted. This means that, if the model represents the field of an
elementary particle, relations (56) indicate that massless charged
particle cannot exist. This result is in agreement with experimental
known results. On the other hand, (56) allows for the existence of
$m=0$ particles, for which ,$$\mu=\frac{1} {2} \frac{\epsilon^{2}}
{a}~~~,~~~ \eqno{(57)}$$ i.e. {\bf particles whose masses are
totally electromagnetic in origin} [15]. If the model represents an
elementary charge ( an electron or a positron), and assuming that
$M$ is the mass of the electron in gm and $Q$ is its charge in
e.s.u. then
$$\mu=\frac{GM} {c^{2}}~~~ ,~~~\epsilon=\frac{c^{\frac{1} {2}} Q}
{c^{2}}~~~,~~~ \eqno{(58)}$$ where $c$ is the speed of light, and
$G$ is the gravitational constant. Substituting from (58) into (57),
we get:
$$2a=\frac{Q^{2}} {c^{2} M} ~~~.~~~ \eqno{(59)}$$ Substituting the
numerical values [16] of $Q$, $M$ and $c$, we get from
(59),$$2a=2.81794\times 10^{-13} cm$$
 this length coincides with
the classical radius of the electron. So, in general $2a$ can be
identified as the radius of a particle whose mass is electromagnetic
in origin.

   (v) It may be of interest to interpret (56) using
classical electrodynamics. A charged particle of radius $2a$ and
charge $Q$ will have a self potential energy (cf. [17] ):
$$\left.
\begin{array}{llll}
\xi&=&\frac{Q^{2}} {4a} &(in ~~ cm ~~  gm~~   sec~~  units)~~,\vspace{2mm}\\
\xi&=&\frac{\epsilon^{2}} {4a}&(in~~ relativistic~~ units)~~.\vspace{2mm}\\
\end{array}
\hspace{1cm} \right\}\eqno{(60)}$$ Now, (56) can be interpreted in
view of (60) as follows: apart from the factor 2, the gravitational
mass of any charged particle will be increased by an amount equal to
the self potential energy of the field produced by the charge.
     An interpretation similar to the above one has been obtained by
Florides [18], by applying Moller's theory on energy [19] to the
well known Riessner-Nordstrom solution  ; and then studying the
motion of a neutral particle in the field of a central charged body.
From the metric point of view, one of the differences between the
present work and that of Florides [19] is that we have obtained a
solution whose associated Riemannian space has the metric of the
Schwarschild exterior form, using which we get the result.

\section{Conclusion}

   The constants of integration have been fixed using certain
boundary conditions. The comparison with classical known results has
given rise to the identification of the constant $ m $ as the
gravitational mass of the body, the constant $\epsilon$ as its
geometrical charge, and the constant $a$ as half its radius. The
appearance of the two constants ~~$m$~~,
and~~$\alpha(=\frac{\epsilon} {a^{\frac{1} {2}}})$ in both
electromagnetic and gravitational quantities shows the interaction
between the two fields.

    The solution allows for the existence of a charged particle
whose total mass is electromagnetic in origin, and independent of
the charge sign. On the other hand, the same solution prevents the
existence of massless charged particles. In general, the solution
shows that the mass of any body, with net electric charge, is made
of two contributions: One is its gravitational mass and the other is
due to its net electric charge.

    Finally,  it is of interest to note that the metric of the
Riemannian space is not sufficient to give a complete indication of
the physical content of the field. It is more convenient to use the
procedure known as the {\it type analysis} in order to get more
insight on the physical fields that a certain space is capable of
representing.
\newpage
\section*{REFERENCES}
{[1] R.Adler, M.Bazin, and M.Schiffer,{\it Introduction to
General Relativity}

(McGraw-Hill 1975) 2nd edition, p.486.}
\\
{[2] F.I.Cooperstock, and V.De La Cruz, Gen.Relativ.Gravit.{\bf 9}
, 835 (1978).} \\
{[3] F.I.Mikhail, and M.I.Wanas, Proc.R.Soc.London
{\bf A 356},471 (1977).} \\
{[4] M.I.Wanas, Nuovo Cimento {\bf 66B},145
 (1981).} \\
{[5] M.I.Wanas, Int.J.Theor.Phys. {\bf 24}, 639 (1985).} \\
{[6] F.I.Mikhail, M.I.Wanas,and A.M.Eid, Astrophys. Space Sci.{\bf
228} , 221 (1995). } \\
{[7] A.Einstein, Math. Annal. {\bf 102}, 685 (1930).}\\
{[8] H.P.Robertson, Ann.Math.Princeton(2),{\bf 33},496 (1932).} \\
{[9] F.I.Mikhail, Ph.D. Thesis, London University (1952).}\\
{[10] N.L.Youssef and A.M.Sidahmed (2006) gr-qc/0604111. }\\
 {[11] F.I.Mikhail
,and M.I.Wanas, Int.J.Theor.Phys.{\bf 20}, 671 (1981).} \\
{[12] M.I.Wanas, Astrophys. Space Sci.{\bf 154} , 165 (1989). }\\
{[13] M.I.Wanas (2001) Proceedings of the 11th National Conference
on

{\it Finsler, Lagrange and Hamilton Geometries}, (Bacau, 2000),

Cercet.Stiin.Ser.Mat 10:,297-309,2001, eds. V.Blanuta and Gh.Neagu;
gr-qc/0209050 }
\\ {[14] M.I.Wanas Astrophys. Space Sci.{\bf
127} , 21 (1986) }  \\
{[15] M.I.Wanas (1987)  ICTP preprint IC{/}87{/}397.  }\\
 {[16] C.W.Allen, {\it Astrophysical
Quantities} (London 1973) 3rd edition,p.14, 15.} \\
{[17] L.D.Landau ,and E.M.Lifshitz,{\it The Classical Theory of
Fields} (Pergamon 1975)

4th  edition, P.89.} \\
{[18] P.S.Florides,
Proc.Camb.Phil.Soc.{\bf 58}, 110 (1962).} \\
{[19] P.S.Florides,
Proc.Camb.Phil.Soc.{\bf 58}, 102 (1962).} \\
\end{document}